\documentclass[article,twocolumn,amsmath,floats,showpacs,nofootinbib]{revtex4}
\usepackage{graphicx}
\usepackage{textcomp}
\usepackage{hyperref}
\hypersetup{colorlinks=true}

\begin{document}
\input epsf

\title{Point-contact electron-phonon interaction function in tantalum}

\author{N. L. Bobrov, L. F. Rybal'chenko, V. V. Fisun, and I. K. Yanson}
\affiliation{B.I.~Verkin Institute for Low Temperature Physics and
Engineering, of the National Academy of Sciences
of Ukraine, prospekt Lenina, 47, Kharkov 61103, Ukraine
Email address: bobrov@ilt.kharkov.ua}
\published {(\href{http://fntr.ilt.kharkov.ua/fnt/pdf/13/13-6/f13-0611r.pdf}{Fiz. Nizk. Temp.}, \textbf{13}, 611 (1987)); (Sov. J. Low Temp. Phys., \textbf{12}, 344 (1987)}
\date{\today}

\begin{abstract}Tantalum is studied by the method of point-contact (PC) spectroscopy: the parameter ${{\lambda }_{pc}}$ and the absolute intensity of the PC EPI function are determined, and the form of the EPI function is determined more accurately. Both homocontacts $Ta-Ta$ and heterocontacts $Ta-Cu$ and $Ta-Au$ were studied. It was found that the contributions of copper and gold to the spectrum of the heterocontact are not noticeable, though the forms of the PC spectra of $Ta$ in homo- and heterocontacts are different. The intensities of the spectra of $Ta-Ta$ homocontacts and $Ta-Cu$ heterocontacts turned out to be close to one another. Simple expressions, enabling numerical calculations of the PC characteristics for geometrically symmetric heterocontacts in the approximation of a spherical Fermi surface, are derived, based on the theory. It is shown that the use of the free-electron approximation in the numerical estimates leads to an error. It is proposed that the effective mass approximation be used for such estimates.

\pacs{71.38.-k, 73.40.Jn, 74.25.Kc, 74.45.+c,}
\end{abstract}

\maketitle

\section{INTRODUCTION}
Point-contact (PC) spectroscopy of the electron- phonon interaction (EPI) in normal metals \cite{Yanson1}  enables direct measurement of the spectral function of EPI under the condition that the inelastic electron mean-free path length ${{l}_{\varepsilon }}$ is greater than the dimensions of the contact and the temperature is quite low ($kT\ll \hbar {{\omega }_{\max }}$, where ${{\omega }_{\max }}$ is the limiting phonon frequency).

The purpose of this work is to obtain quantitative information about the EPI in tantalum - determine the absolute intensity of the PC EPI function ${{g}_{_{pc}}}(\omega )$ and the parameter ${{\lambda }_{pc}}$ and also to determine more accurately the form of the PC EPI function. A new method was employed to prepare pure $Ta-Ta$ homocontacts and $Ta-Cu$ and $Ta-Au$ heterocontacts, which enabled us to obtain high-intensity spectra. The value of the EEP parameter ${{\lambda }_{pc}}$ in a pure homocontact is 6.5 times greater than the value obtained in the preceding work \cite{Rybal'chenko}. It is demonstrated that quantitative information can be obtained about the EPI of tantalum in $Ta-Cu$ heterocontacts. It was found that the contributions of copper and gold to the PC spectrum of the heterocontact are not noticeable though the form of the PC spectra of $Ta$ is different in homo- and heterocontacts. The reasons for these phenomena are established. It is shown that heterocontacts enable the determination of the ratio of the Fermi velocities of the corresponding electrodes.

\section{EXPERIMENTAL PROCEDURE}

A tantalum single-crystal with the resistance ratio ${{{\rho }_{300}}}/{{{\rho }_{5}}\sim{\ }20}\;$, a copper single-crystal with the resistance ratio ${{{\rho }_{300}}}/{{{\rho }_{4.2}}\sim{\ }1000}\;$ and a gold polycrystal with approximately the same value of ${{{\rho }_{300}}}/{{{\rho }_{4.2}}}\;$ were employed as the material for the electrodes. The electron momentum mean-free path length in tantalum at liquid-helium temperature, determined by the impurities, equals about 900 {\AA} for our samples. In the estimate  $\rho l=0.59\cdot {{10}^{-11}}\Omega \cdot c{{m}^{2}}$ \cite{Ryazanov} and ${{\rho }_{273}}=12.6\cdot {{10}^{-6}}\Omega \cdot cm$\ \cite{Startsev} were used. Electrodes with dimensions of  $1.5\times 1.5\times 10\ m{{m}^{3}}$ were cut out by the electric-spark method and subjected to chemical polishing in a mixture of concentrated acids. For tantalum, the mixture consisted of $\text{HF}\ \text{:}\ \text{HN}{{\text{O}}_{\text{3}}}\text{:}\ \text{HCl}{{\text{O}}_{\text{4}}}$, taken in equal volume ratios, while for copper the mixture consisted of $\text{HN}{{\text{O}}_{\text{3}}}\ \text{:}\ {{\text{H}}_{\text{3}}}\text{P}{{\text{O}}_{\text{4}}}\ \text{:}\ \text{C}{{\text{H}}_{\text{3}}}\text{COOH}$ in the volume ratio 2:1:1. The gold electrodes were chemically treated in aqua regia. The electrodes were then flushed in distilled water, dried, and mounted in a clamping device.

The contacts were created by the modified shear method \cite{Yanson1}, and in so doing, in order to obtain stable pure contacts, the following procedure was employed: A reference current equal to $10-50\ \mu A$ was fixed with completely shunted samples, after which the shunts were switched off and a contact was obtained with a resistance from several hundreds to several thousands of ohms, most often with a dependence R(V) of a nonmetallic type. After this the current was increased in steps with the help of a resistance box connected in series with the point contact. The initial resistance of the box was much greater than the resistance of the point contact. The voltage on the point contact in this case, as a rule, varied insignificantly and equaled $500\pm 200\ mV$. When the required resistance was achieved the contact was held under that current for several minutes. The contacts prepared in this manner were much cleaner than those prepared by the standard procedure \cite{Yanson1} and were distinguished by high mechanical and electric stability.

We note that the quality of the spectra was significantly higher than those which we obtained previously\cite{Rybal'chenko}. At the same time, together with a general increase in the intensity of the spectrum, the relative intensity of the   peak increases significantly. This can be explained by the decrease in the degree of deformation of the material in the region of the contact \cite{Bostok} when it is prepared. The critical fields required for suppressing superconductivity also dropped by a factor of 3-5. With the new method for forming the contact with bias voltages of several hundreds of millivolts, the contact was in the normal regime. The temperature at the center of the contact reaches the softening point of the metal and, unlike impulsive breakdown, is maintained at this level for quite a long time, so as to order the crystal lattice of the metal in the region near the contact. Additional purification of this region apparently also occurs at the same time by the electric transport method. We note that for electroforming the heating in the $Ta-Cu$ heterocontacts is not symmetric: only the tantalum edge is heated. This is linked with the fact that for high bias voltages the flight of the electrons from the copper side remains close to ballistic. This is indirectly confirmed by the change in the differential resistance - for $Cu-Cu$ homocontacts with bias voltages of the order of 300-350 $meV$ it increases by 20-40\%  compared with $eV\sim30\ meV$, while for $Ta-Cu$ heterocontacts and $Ta-Ta$ homocontacts it more than doubles for such bias voltages.

Since the short-circuiting begins most often with a resistance of several kiloohms, while the final resistances of the point contact equal, as a rule, not more than several tens of ohms, the shunting of the short circuit by the tunneling gap of the reference oxide by the comparison resistance is eliminated. Multiple contacts are also unlikely to form under conditions of electroforming. Another advantage of this method is the possibility of "growing" short-circuits with the required resistance with an accuracy of several percent. The same method was also employed to prepare $Ta-Au$ point contacts, i.e., the method is in principle also applicable to metals which are mutually soluble (copper and tantalum are not mutually soluble in the solid and liquid phases).

Some of the $Ta-Ta$, $Ta-Cu$, and $Ta-Au$ point contacts were prepared by the standard shear method. The optimal resistances of the point contacts, prepared by electro-forming, turned out to be several times higher than the corresponding values for point contacts prepared by the standard shear method. This could be attributable to the difference in the form of the short-circuits formed: In the shear method the short-circuit apparently forms along cracks in the reference oxide, and in this case its form is described best by the model of a prolate ellipse. For electroforming, the form of the short circuit is probably closer to a circular opening. Measurements were performed in an intermediate cryostat with a capillary, structurally analogous to that described in Ref.\cite{Engen}, in the temperature range 1.4-6 K. For measurements of the PC characteristics in the normal state the superconductivity was destroyed either with a magnetic field from a superconducting solenoid or by raising the temperature above ${{T}_{c}}$ for tantalum.

\section{EXPERIMENTAL RESULTS AND DISCUSSION}

\begin{figure}[]
\includegraphics[width=8cm,angle=0]{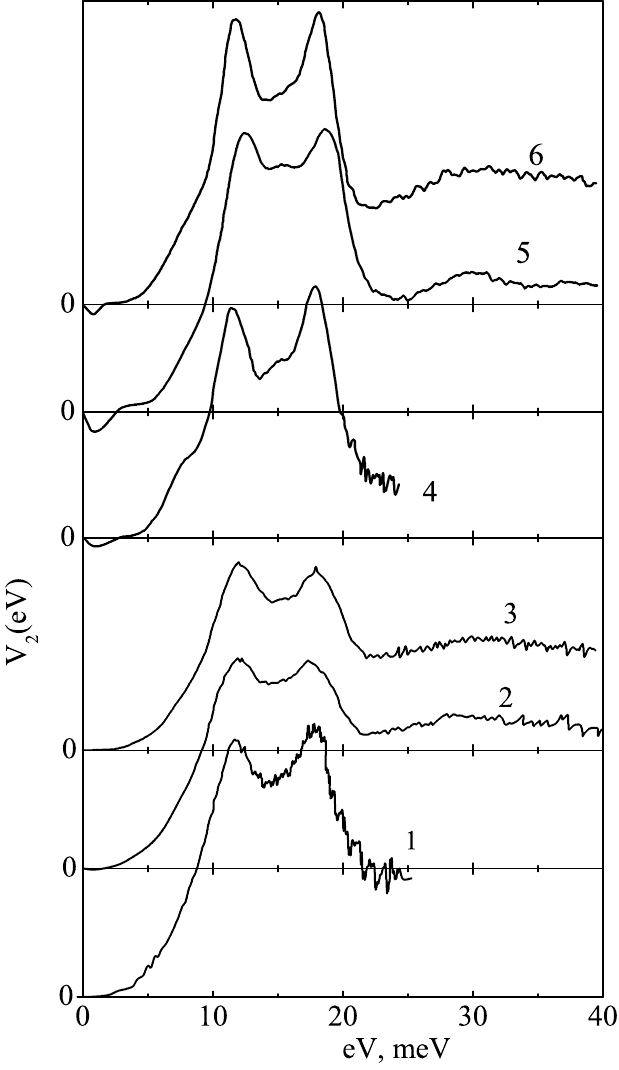}
\caption[]{PC EPI spectra ${{V}_{2}}\sim{\ }{{{d}^{2}}V}/{d{{I}^{2}}}\;$ in the N state under conditions of electroforming:\\
1) $Ta-Ta$, $R=50\ \Omega $, ${{V}_{1}}(0)=300\ \mu V$, $V_{2}^{\max }=0.355\ \mu V$, $T=4.8\ K$, $H=0$;\\
2) $Ta-Ta$, $R=78\ \Omega $, ${{V}_{1}}(0)=336\ \mu V$, $V_{2}^{\max }=0.401\ \mu V$, $T=4.6\ K$, $H=0$;\\
3) $Ta-Ta$, $R=64\ \Omega $, ${{V}_{1}}(0)=347\ \mu V$, $V_{2}^{\max }=0.532\ \mu V$, $T=4.6\ K$, $H=0$;\\
4) $Ta-Cu$, $R=80\ \Omega $, ${{V}_{1}}(0)=492\ \mu V$, $V_{2}^{\max }=0.787\ \mu V$, $T=1.88\ K$, $H=3\ kOe$;\\
5) $Ta-Cu$, $R=73\ \Omega $, ${{V}_{1}}(0)=612\ \mu V$, $V_{2}^{\max }=0.594\ \mu V$, $T=1.65\ K$, $H=5.6\ kOe$;\\
6) $Ta-Cu$, $R=73\ \Omega $, ${{V}_{1}}(0)=497\ \mu V$, $V_{2}^{\max }=0.874\ \mu V$, $T=1.72\ K$, $H=2.4\ kOe$.}
\label{Fig1}
\end{figure}

For a detailed investigation the point contacts were selected beforehand according to quality criteria described in Ref.\cite{Yanson1} Because the electron energy mean-free path length at energies close to the Debye energy is quite short, the best results were obtained for comparatively high-resistance samples, whose resistance fell into the range from 20 to $100\ \Omega$. The typical PC spectra ${{V}_{2}}(eV)\sim{\ }{{{d}^{2}}V}/{d{{I}^{2}}}\;$ are presented in Fig. \ref{Fig1}.

\begin{figure}[]
\includegraphics[width=5cm,angle=0]{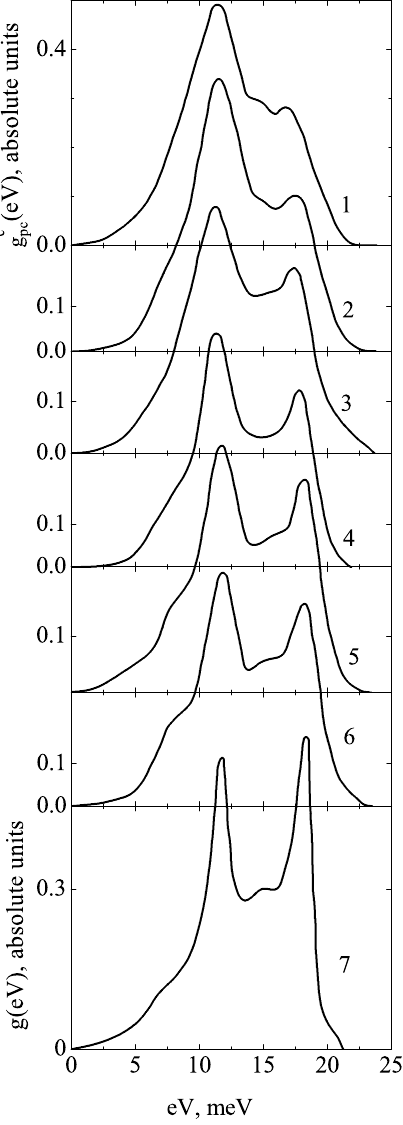}
\caption[]{PC EPI function in Ta:\\
1) $Ta-Ta$, from 2 (Fig. \ref{Fig1}), $\lambda _{pc}^{f}=0.764$, $g_{pc}^{c\ \max }=0.500$;\\
2) $Ta-Ta$, from 3 (Fig. \ref{Fig1}),$\lambda _{pc}^{f}=0.882$, $g_{pc}^{c\ \max }=0.622$;\\
3) $Ta-Ta$, from 1 (Fig. \ref{Fig1}), $\lambda _{pc}^{f}=0.768$, $g_{pc}^{c\ \max }=0.476$;\\
4) $Ta-Cu$; from 6 (Fig. \ref{Fig1}), $\lambda _{pc}^{f}=0.770$, $g_{pc}^{c\ \max }=0.570$;\\
5) $Ta-Cu$, spectrum not presented, $\lambda _{pc}^{f}=0.696$, \\$g_{pc}^{c\ \max }=0.446$;\\
6) $Ta-Cu$, from 4 (Fig. \ref{Fig1}),$\lambda _{pc}^{f}=0.820$, $g_{pc}^{c\ \max }=0.554$;\\
7) from\cite{Shen2}, $\lambda =0.69,\quad {{g}_{\max }}=0.594$.}
\label{Fig2}
\end{figure}
The curves 1-3 refer to $Ta-Ta$ homocontacts, while the curves 4-6 refer to $Ta-Cu$ contacts. The orientation of the contact axis relative to the axis of the crystal lattice of tantalum was not monitored and was random. For this reason the position of the phonon peaks could have varied along the energy scale from 11.5 to 12.5 $meV$ and from 17.5 to 18 $meV$. In many spectra a soft mode at 7-7.5 $meV$ as well as a break near 3 $meV$ are clearly visible. In spectra measured at low temperature, a feature in the region of 15 $meV$ is clearly visible. Figure \ref{Fig2} shows the PC EPI functions of tantalum, reconstructed from these spectra. The absolute values of the EPI functions $q{{g}_{pc}}(\omega )$ and the parameter ${{\lambda }_{pc}}$ were calculated in the free-electron model for contacts in the form of a circular opening \cite{Kulik}:
\begin{equation} \label{eq__1}
{{\lambda }_{pc}}=2\int{{{g}_{pc}}(\omega ){d\omega }/{\omega }\;}
\end{equation}
\begin{equation} \label{eq__2}
\begin{matrix}
  {{g}_{pc}}(\omega )=\frac{3\hbar }{2\sqrt{2}e}{{v}_{F}}\frac{{{{\tilde{V}}}_{2}}}{V_{{{1.0}^{d}}}^{2}}= \\
  =0.6988\cdot {{10}^{-8}}v\ [cm/\sec ]\frac{{{{\tilde{V}}}_{2}}[V]}{V_{1.0}^{2}[V]d[nm]}. \\
\end{matrix}
\end{equation}
Here the voltage of the second harmonic of the modulating signal ${{V}_{2}}(eV)\sim{\ }{{{d}^{2}}V}/{d{{I}^{2}}}\;$; ${{V}_{1.0}}$ is the value of the modulating voltage with zero bias; $d={{\left( {16\rho l}/{3\pi {{R}_{0}}}\; \right)}^{{1}/{2}\;}}$ is the diameter of the contact;
\begin{equation} \label{eq__3}
\rho l={{p}_{F}}/n{{e}^{2}}=\frac{3{{\pi }^{2}}\hbar }{k_{F}^{2}{{e}^{2}}}=\frac{1.66\cdot {{10}^{4}}}{{{\left\{ {{k}_{F}}\left( c{{m}^{-1}} \right) \right\}}^{2}}}\ \left[ \Omega \cdot c{{m}^{2}} \right];
\end{equation}
\begin{equation} \label{eq__4}
d=\frac{4}{e{{k}_{F}}}{{\left( \frac{\pi \hbar }{{{R}_{0}}} \right)}^{{1}/{2}\;}}=44.49\frac{{{10}^{8}}{{\left( R\left[ \Omega  \right] \right)}^{-1/2}}}{{{k}_{F}}\left[ c{{m}^{-1}} \right]}\left[ nm \right].
\end{equation}
In the free-electron model \cite{Harrison}\\
\begin{equation} \label{eq__5}
{{k}_{F}}={{\left( {3{{\pi }^{2}}z}/{\Omega }\; \right)}^{{1}/{3}\;}},
\end{equation}
where $z$ is the number of conduction electrons per unit cell; $\Omega $ is the volume of the unit cell. For a $bbc$ lattice
\begin{equation} \label{eq__6}
\Omega ={{{a}^{3}}}/{2}\;,
\end{equation}
where $a$ is the lattice constant. In the free-electron approximation the true wave functions are approximated by smooth pseudowave functions. The greatest differences are observed in this case in the region of the atomic core which in simple metals is small and occupies about 10\% of the volume. In transport phenomena, in particular, electric conduction, the free-electron approximation in metals such as, for example, copper and gold, "works" very well. In the VA subgroup for $V$, $Nb$, and $Ta$ over filled shells with the configuration of argon, krypton, and xenon, respectively, each atom has five valence electrons. Because of the small number of electrons filling the d bands, the Fermi level intersects them, and for this reason the band structure of these metals is very complicated near the Fermi surface. All metals of the subgroup are uncompensated with a total number of carriers equal to one hole per atom \cite{Startsev}. Therefore, for tantalum, like for niobium in Ref.\cite{Ajhcroft}, we take \emph{z}=1, in the free-electron approximation. At the same time, taking into account the fact that a=3.296 {\AA} \cite{Startsev}, we obtain ${{k}_{F}}=1.183\cdot {{10}^{8}}\ c{{m}^{-1}}$, ${{V}_{F}}=1.396\cdot {{10}^{8}}\,cm/\sec $, $\rho l=0.834\cdot {{10}^{-11}}\ \Omega \cdot c{{m}^{2}}$, $d=37.6{{\left( R\left[ \Omega  \right] \right)}^{{-1}/{2}\;}}\ \left[ nm \right]$. In this approximation the maximum value obtained in our experiments on homocontacts ${{\lambda }_{pc}}=0.88$, which is greater than the EPI parameter $\lambda =0.73$, measured with the help of the tunnel effect \cite{Wolf}, but is less than $\lambda =1.17$, calculated from first principles \cite{Pishche}. We shall evaluate the expected contribution of copper to the spectrum of the heterocontact $Ta-Cu$ for the symmetric geometry. It follows from Ref.\cite{Pishche} that
\begin{equation} \label{eq__7}
L={{{\left( \frac{1}{R}\frac{dR}{dv} \right)}_{1}}}\Big/{{{\left( \frac{1}{R}\frac{dR}{dv} \right)}_{2}}=\frac{{{v}_{{{F}_{2}}}}}{{{v}_{{{F}_{1}}}}}}\;{{\left( \frac{{{p}_{{{F}_{2}}}}}{{{p}_{{{F}_{1}}}}} \right)}^{2}}\frac{g_{pc}^{(1)}}{g_{pc}^{(2)}}.
\end{equation}
Since in Ref.\cite{Shekhter} the analogous relation was obtained for model metals, which have PC EPI functions of the same form and absolute intensity but with different Fermi velocities and momenta, the cofactor ${g_{pc}^{(1)}}/{g_{pc}^{(2)}}\;$ does not occur there. Equation (\ref{eq__7}) does not agree with the expression which would be obtained in an analysis of the relative intensity of the spectra of homocontacts with the same diameter. The additional factor ${{({{p}_{{{F}_{1}}}}/{{p}_{{{F}_{1}}}})}^{2}}$  arises because the point contact form factors of metals with different momenta in heterocontacts do not equal one another and correspondingly differ from the form factors of the homocontact. In the metal with high ${{p}_{F}}$, the relative phase volume of states filled in a nonequilibrium manner is smaller because some of the electron trajectories are reflected from the interface. As shown in Ref.\cite{Shekhter},
${{\left( {{{p}_{{{F}_{1}}}}}/{{{p}_{{{F}_{2}}}}}\; \right)}^{2}}=\left\langle {{{K}_{1}}}/{{{K}_{2}}}\; \right\rangle $ where $\left\langle {{K}_{2}} \right\rangle $ is the angle-averaged form factor of the s-th metal in the heterocontact. The formula (\ref{eq__7}) was obtained for a heterocontact formed by dirty metals, but, since, according to Ref.\cite{Shekhter}, the ratio of the intensities of the spectra of the metals is independent of the electron momentum mean-free path length and remains correct in the regime close to the ballistic regime\footnote {Private remark by R. L. Shekhter}, we shall use it in what follows, making the assumption that our contacts are clean.

Our maximum absolute intensity of the PC EPI function of tantalum in homocontacts in the free-electron approximation was $g_{pc}^{Ta}{{\,}_{\max }}=0.621$; for copper in the same approximation $g_{pc}^{Cu}{{\,}_{\max }}=0.241$ [13], $v_{F}^{Cu}=1.57\cdot {{10}^{8}}\,cm/\sec $, $k_{F}^{Cu}=1.36\cdot {{10}^{8}}\,c{{m}^{-1}}$, and then
\begin{equation} \label{eq__8}
{{L}_{c}}={{\left[\left(\frac{1}{R}\frac{dR}{dv}\right)_{Cu}^{\max}\Big/\left(\frac{1}{R}\frac{dR}{dv}\right)_{Ta}^{\max}\right]}_{c}}\simeq0.255.
\end{equation}
We shall go beyond the free-electron approximation and determine this ratio employing the electronic parameters of tantalum and copper, which were determined based on the experimental data. Taking into account the fact that
\begin{equation} \label{eq__9}
\rho l=3{{\left( 2N(0){{v}_{F}}{{e}^{2}} \right)}^{-1}};\ \ \gamma ={}^{2}/{}_{3}{{\pi }^{2}}k_{B}^{2}N(0)\left( 1+\lambda  \right),
\end{equation}
we obtain
\begin{equation} \label{eq__10}
\begin{matrix}
  \rho l{{v}_{F}}=\frac{3}{2N(0){{e}^{3}}}={{\left( \frac{\pi {{k}_{B}}}{e} \right)}^{2}}\frac{1+\lambda }{\gamma }= \\
  =0.7347\frac{\left( 1+\lambda  \right)}{\gamma \left[ erg\cdot c{{m}^{-3}}\cdot {{K}^{-2}} \right]}\left[ \Omega \cdot c{{m}^{3}}\cdot {{\sec }^{-1}} \right] \\
\end{matrix}.
\end{equation}
Here N(0) is the single-spin unrenormalized electron density of states at the Fermi level and $\gamma $ is the coefficient of electronic thermal conductivity. Using ${{\gamma }_{Ta}}=4.36\,mJ\cdot mol{{e}^{-1}}{{K}^{-2}}=4\cdot {{10}^{3}}erg\cdot c{{m}^{-3}}{{K}^{-2}}$\cite{Startsev}, $\rho {{l}_{Ta}}=0.59\cdot {{10}^{-11}}\Omega \cdot c{{m}^{2}}$\cite{Ryazanov}, $\lambda =0.65$, we obtain
\begin{equation} \label{eq__11}
\begin{matrix}
  {{\left( \rho l{{v}_{F}} \right)}_{Ta}}=303\cdot {{10}^{-6}}\Omega \cdot c{{m}^{3}}\cdot {{\sec }^{-1}};\  \\
  v_{F}^{Ta}=0.51\cdot {{10}^{+8}}cm/\sec.  \\
\end{matrix}
\end{equation}
This quantity is identical to the value of $\left\langle v_{F}^{2} \right\rangle _{Nb}^{{1}/{2}\;}$, determined in Ref.\cite{Mattheiss}, and is quite close to $\left\langle v_{F}^{2} \right\rangle _{Nb}^{{1}/{2}\;}=0.63\cdot {{10}^{8}}cm/\sec $ from Ref.\cite{Alekseevskii}, which is the true consequence of the fact that niobium and tantalum are electronic analogs; their Fermi energies are virtually identical, and the Fermi surfaces are very close \cite{Mattheiss}.

Because of the very complicated Fermi surface in tantalum the use of the formula \eqref{eq__3} for determining ${{k}_{F}}$ will lead to a large error. To evaluate $k_{{{F}_{e}}}^{Ta}$ we shall employ our value of $v_{{{F}_{e}}}^{Ta}$ and the results of Ref.\cite{Halloran}, in which the ratio of the effective electron mass, averaged over the Fermi surface, to the free-electron value  ${{{m}^{*}}}/{{{m}_{0}}=1.85}\;$ was determined from the de Haas-van Alphen effect. Then
\begin{equation} \label{eq__12}
k_{{{F}_{e}}}^{Ta}={{m}^{*}}v_{{{F}_{e}}}^{Ta}{{\hbar }^{-1}}=0.82\cdot {{10}^{-8}}c{{m}^{-1}}.
\end{equation}
The Fermi surface of copper is not very different from a spherical surface, so that the formula \eqref{eq__3} and the experimental value $\rho {{l}_{Cu}}0.53\cdot {{10}^{-11}}\ \Omega \cdot c{{m}^{2}}$ can be employed without making any special error \cite{Gniwek}. Substituting the numerical value of $\rho {{l}_{Cu}}$ into the formula \eqref{eq__3} we obtain
\begin{equation} \label{eq__13}
k_{{{F}_{e}}}^{Cu}=1.48\cdot {{10}^{-8}}c{{m}^{-1}}.
\end{equation}
We can now find the absolute intensity of the PC EPI function ${{g}_{pc}}$ and the parameter ${{\lambda }_{pc}}$ using the experimental data. The transfer factor from the free-electron to the experimental values for tantalum equals
\begin{equation}
\label{eq__14}
\begin{matrix}
  {{k}_{t}}=\lambda _{pc}^{f}/\lambda _{pc}^{e}=g_{pc}^{f}/g_{pc}^{e}=v_{F}^{f}{{d}^{e}}/v_{F}^{e}{{d}^{f}}= \\
  =v_{F}^{f}k_{F}^{f}/v_{F}^{e}k_{F}^{e}=3.87. \\
\end{matrix}
\end{equation}
From here ${{\lambda }_{pc}}$=0.227. The corresponding formulas for determining the diameter of tantalum and copper homocontacts have the form
\begin{equation} \label{eq__15}
\begin{matrix}
  d_{Ta}^{e}=54.26{{\left( R\left[ \Omega  \right] \right)}^{{-1}/{2}\;}}\left[ nm \right]; \\
  d_{Cu}^{e}=30{{\left( R\left[ \Omega  \right] \right)}^{{-1}/{2}\;}}\left[ nm \right]. \\
\end{matrix}
\end{equation}
The contribution of copper to the spectrum of the heterocontact, using the experimental electronic parameter s, is less than in the free-electron approximation. It follows from the formulas \eqref{eq__7} and \eqref{eq__14} that
\begin{equation} \label{eq__16}
{{{L}_{c}}}/{{{L}_{e}}={{\left( {k_{F}^{Ta}}/{k_{F}^{Cu}}\; \right)}_{f}}}\;{{\left( {k_{F}^{Cu}}/{k_{F}^{Ta}}\; \right)}_{e}}=1.57.
\end{equation}
Here the expected relative contribution of copper to the spectrum of the heterocontact, using the electronic parameters of tantalum and copper determined from the experimental data, is given by
\begin{equation} \label{eq__17}
{{L}_{e}}=0.162.
\end{equation}
For what follows, we note that the maximum absolute intensity of the PC EPI function of copper with the experimental value for ${{k}_{F}}$ (see the formula \eqref{eq__14}) equals
\begin{equation} \label{eq__18}
g_{pc\max }^{Cu}=0.262.
\end{equation}
The point-contact EPI function ${{g}_{pc}}(\omega )$ is related to Eliashberg's thermodynamic function $g(\omega )$, determined in experiments on the elastic tunneling effect in superconductors and the transport EPI function ${{g}_{tr}}(\omega )$, which is determined from precision measurements of the temperature dependence of the resistivity, owing to scattering by phonons. Only the factor in the integrand - the form factor, determined by the scattering angle - changes. The role of electron scattering by large angles increases in the sequence $g\to {{g}_{tr}}\to {{g}_{pc}}$. The transport EPI parameter satisfies the relation
\begin{equation} \label{eq__19}
{{\lambda }_{tr}}=\left( {\hbar }/{2\pi }\; \right)\left( {{{v}_{F}}}/{\rho l{{k}_{B}}}\; \right)\partial \rho /\partial T,
\end{equation}
where $\partial \rho /\partial T$ is the slope of the temperature dependence of the resistivity at high temperatures$\left( T>{{\Theta }_{D}} \right)$. Using the experimental values for $\rho l$ and ${{v}_{F}}$ for tantalum we obtain ${{\lambda }_{tr}}=0.456$ \cite{Yanson1}. If, however, $\rho l$ and ${{v}_{F}}$, calculated in the free-electron approximation, are employed, then we obtain ${{\lambda }_{tr}}=0.854$. Since the transport EPI function ${{g}_{tr}}$ occupies an intermediate position between the PC EPI function ${{g}_{pc}}$ and Eliashberg's thermodynamic function $g$, with the use of the experimental electronic parameters, the relation ${{\lambda }_{pc}}=0.227<{{\lambda }_{tr}}=0.456<\lambda =0.73$ holds.
We shall now determine the absolute intensity of the partial spectrum of tantalum in a symmetric heterocontact. The resistance of a heterocontact in the presence of a $\delta $ function barrier at the boundary between the metals equals \cite{Shekhter}
\begin{equation} \label{eq__20}
R_{het}^{-1}=\frac{{{e}^{2}}S{{S}_{F}}}{{{\left( 2\pi \hbar  \right)}^{3}}}{{\left\langle \alpha D(\alpha ) \right\rangle }_{\begin{smallmatrix}
 1,2 \\
 {{v}_{z}}>0
\end{smallmatrix}}}.
\end{equation}
Here $\left\langle ... \right\rangle {{v}_{z}}>0$ denotes averaging over the Fermi surfaces of metals 1 and 2, respectively, under the condition ${{v}_{z}}>0$; S is the area of the contact; ${{S}_{F}}$ is the area of the Fermi surface; $\alpha ={{{v}_{z}}}/{{{v}_{F}}=\cos \theta }\;$; and, $D$ is the transmission coefficient of the boundary. The quantity $R_{het}^{-1}$ is independent of the metal over which the averaging is performed, and it is independent of the electron dispersion law, i.e.,
\begin{equation} \label{eq__21}
{{\left\{ {{S}_{F}}\left\langle \alpha D(\alpha ) \right\rangle  \right\}}_{1}}={{\left\{ {{S}_{F}}\left\langle \alpha D(\alpha ) \right\rangle  \right\}}_{2}}.
\end{equation}
Taking into account the fact that
\begin{equation} \label{eq__22}
\frac{{{\left( 2\pi \hbar  \right)}^{3}}}{{{e}^{2}}S{{S}_{F}}{{\left\langle \alpha  \right\rangle }_{{{v}_{z}}>0}}}=\frac{16\pi \hbar }{{{e}^{2}}k_{F}^{2}{{d}^{2}}}={{R}_{0}},
\end{equation}
where ${{R}_{0}}$ is the resistance of the homocontact, we obtain ${{R}_{het}}={{R}_{0}}\left( {{{\left\langle \alpha  \right\rangle }_{{{v}_{z}}>0}}}/{{{\left\langle \alpha D(\alpha ) \right\rangle }_{{{v}_{z}}>0}}}\; \right)$. For a spherical Fermi surface
\begin{equation} \label{eq__23}
\begin{matrix}
  {{\left\langle \alpha  \right\rangle }_{{{v}_{z}}>0}}=1/2;\quad {{\left\langle \alpha D(\alpha ) \right\rangle }_{{{v}_{z}}>0}}=\int\limits_{0}^{1}{\alpha D(\alpha )}d\alpha ; \\
  R_{het}^{-1}=2R_{0}^{-1}\int\limits_{0}^{1}{\alpha D(\alpha )}d\alpha.  \\
\end{matrix}
\end{equation}
The expression for the form factors of the metals in heterocontacts was calculated in Ref.\cite{Shekhter}:
\begin{equation} \label{eq__24}
{{K}_{s}}\left( \mathbf{p},\mathbf{{p}'} \right)=\frac{{{D}_{12}}({{\mathbf{p}}_{\mathbf{s}}}=\mathbf{p}){{D}_{12}}({{\mathbf{p}}_{\mathbf{s}}}=\mathbf{{p}'})}{4\left\langle \alpha D(\alpha ) \right\rangle }{{K}_{0s}}\left( \mathbf{p},\mathbf{{p}'} \right).
\end{equation}
Here ${{K}_{s}}$ denotes ${{K}_{Ta}}$ or ${{K}_{Cu}}$;   ${{K}_{0s}}\left( \mathbf{p},\mathbf{{p}'} \right)$ are the form factors of the pure homocontacts, which are identical for both metals in the clean limit $\left\langle {{K}_{0}} \right\rangle ={1}/{4}\;$ in the model of an opening. For direct contact between the metals we have \cite{Shekhter}
\begin{equation}
\label{eq__25}
D=\frac{4{{v}_{z1}}{{v}_{z2}}}{{{\left( {{v}_{z1}}+{{v}_{z2}} \right)}^{2}}},
\end{equation}
where ${{v}_{z1}}={{v}_{F1}}\cos {{\theta }_{1}}$; ${{v}_{z2}}={{v}_{F2}}\cos {{\theta }_{2}}$; from the law of conservation of momentum $p_\parallel ={{p}_{F1}}\sin {{\theta }_{1}}={{p}_{F2}}\sin {{\theta }_{2}}$. We denote ${{{p}_{F1}}}/{{{p}_{F2}}=b}\;$; ${{{v}_{F1}}}/{{{v}_{F2}}=c}\;$; $\cos {{\theta }_{1}}={{\alpha }_{1}}$; $\cos {{\theta }_{2}}={{\alpha }_{2}}$. We assume for definiteness that $b<1$. As a result we obtain
\begin{equation}
\label{eq__26}
{{\alpha }_{1}}={{b}^{-1}}{{\left( \alpha _{2}^{2}+{{b}^{2}}-1 \right)}^{{1}/{2}\;}};\ {{\alpha }_{2}}=b{{\left( \alpha _{1}^{2}+{{b}^{-2}}-1 \right)}^{{1}/{2}\;}}.
\end{equation}
The transmission coefficients at each edge can be written in the form
\begin{equation}
\label{eq__27}
D\left( {{\alpha }_{1}} \right)=\frac{4b{{\alpha }_{1}}{{\left( \alpha _{1}^{2}+{{b}^{-2}}-1 \right)}^{{1}/{2}\;}}}{c{{\left[ {{\alpha }_{1}}+\left( b/c \right){{\left( \alpha _{1}^{2}+{{b}^{-2}}-1 \right)}^{{1}/{2}\;}} \right]}^{2}}};
\end{equation}
\begin{equation}
\label{eq__28}
D\left( {{\alpha }_{2}} \right)=\frac{4c{{\alpha }_{1}}{{\left( \alpha _{2}^{2}+{{b}^{2}}-1 \right)}^{{1}/{2}\;}}}{b{{\left[ {{\alpha }_{2}}+\left( c/b \right){{\left( \alpha _{2}^{2}+{{b}^{2}}-1 \right)}^{{1}/{2}\;}} \right]}^{2}}}.
\end{equation}
We shall solve the problem from the side of the tantalum edge. In this case
\begin{equation} \label{eq__29}
2{{\left\langle {{\alpha }_{1}}D\left( {{\alpha }_{1}} \right) \right\rangle }_{{{v}_{z}}>0}}=2\int\limits_{0}^{1}{{{\alpha }_{1}}D\left( {{\alpha }_{1}} \right)d{{\alpha }_{1}}}=0.597.
\end{equation}
Here we assume that $b={p_{F}^{Ta}}/{p_{F}^{Cu}=0.554}\;$; $c={v_{F}^{Ta}}/{v_{F}^{Cu}=0.325}\;$. For the diameter of the $Ta-Cu$ heterocontact we have from \eqref{eq__22} and \eqref{eq__23}
\begin{equation} \label{eq__30}
{{d}_{het}}={{d}_{Ta}}{{\left[ 2{{\left\langle {{\alpha }_{1}}D({{\alpha }_{1}}) \right\rangle }_{{{v}_{z}}>0}} \right]}^{{-1}/{2}\;}}=70.2{{\left( R\left[ \Omega  \right] \right)}^{{-1}/{2}\;}}[nm],
\end{equation}
where $d_{Ta}$ is the diameter of the tantalum homocontact with the same resistance (see \eqref{eq__15}). The problem can also be solved from the copper side:
\begin{equation} \label{eq__31}
\begin{matrix}
  2{{\left\langle {{\alpha }_{2}}D({{\alpha }_{2}}) \right\rangle }_{{{v}_{z}}>0}}=2\int\limits_{\sqrt{1-{{b}^{2}}}}^{1}{{{\alpha }_{2}}D({{\alpha }_{2}})}d{{\alpha }_{2}}\equiv  \\
  \equiv 2{{b}^{2}}{{\left\langle {{\alpha }_{1}}D({{\alpha }_{1}}) \right\rangle }_{{{v}_{z}}>0}}=0.1833. \\
\end{matrix}
\end{equation}
Here the integration is performed from $\sqrt{1-{{b}^{2}}}$, because of the fact that for ${{\alpha }_{2}}\ll \sqrt{1-{{b}^{2}}}$ the quantity $D\left( {{\alpha }_{2}} \right)=0$, since total internal reflection of the electrons occurs. The diameter of the heterocontact in this case equals
\begin{equation} \label{eq__32}
\begin{matrix}
  {{d}_{het}}={{d}_{Cu}}{{b}^{-1}}{{\left[ 2{{\left\langle {{\alpha }_{1}}D({{\alpha }_{1}}) \right\rangle }_{{{v}_{z}}>0}} \right]}^{-1/2\;}}= \\
  =70.2{{\left( R\left[ \Omega  \right] \right)}^{-1/2\;}}[nm], \\
\end{matrix}
\end{equation}
i.e., the same result is obtained as in the calculation from the tantalum side. (We note that if the diameter of the tantalum heterocontact is determined from Sharvin's formula
	\[{{d}_{Ta}}={{\left( \frac{16\rho l}{3\pi } \right)}^{{1}/{2}\;}}{{R}^{{-1}/{2}\;}}=31.65{{\left( R\left[ \Omega  \right] \right)}^{{-1}/{2}\;}}[nm],\]
then we obtain
	\[{{d}_{het}}=41.1{{\left( R\left[ \Omega  \right] \right)}^{{-1}/{2}\;}}[nm],\]
i.e., the calculation from the copper side yields a result which is different from the calculation on the tantalum side. The fact that the simple relation \eqref{eq__21} does not hold is apparently linked with the presence of several groups of carriers in the tantalum, and this makes it necessary to generalize the relation \eqref{eq__21} to this case.)
To determine the averaged form factors of tantalum and copper in the heterocontact in the ballistic state, the numerator in the formula \eqref{eq__24} must first be calculated:
 \[{{A}_{S=Ta,}}\ {{B}_{S=Cu}}={{\left\langle {{D}_{12}}({{\mathbf{p}}_{\mathbf{s}}}=\mathbf{p}){{D}_{12}}({{\mathbf{p}}_{\mathbf{s}}}=\mathbf{{p}'}){{K}_{0s}}(\mathbf{p},\mathbf{{p}'}) \right\rangle }_{{{v}_{z}}>0}}\]
In the model of an opening \cite{Shekhter}
\begin{equation} \label{eq__33}
{{K}_{0s}}={{K}_{0}}(\mathbf{p},\mathbf{{p}'})=\frac{\left| {{v}_{z}}{{v}_{z}}^{\prime } \right|\theta \left( -{{v}_{z}}{{v}_{z}}^{\prime } \right)}{\left| {{v}_{z}}\mathbf{{v}'}-{{v}_{z}}\mathbf{v} \right|}
\end{equation}
($\theta $ is the Heaviside function).
Decomposing the numerator of the formula into components and transforming it for a spherical Fermi function we obtain
\begin{equation} \label{eq__34}
A=\frac{32}{\pi }\int\limits_{0}^{1}{dx}\int\limits_{0}^{1}{dy\frac{xD\left( {{\alpha }_{1}}=x \right)D\left( {{\alpha }_{1}}=y \right)K(k)}{{{(n+m)}^{2}}}},\ y>x;
\end{equation}
\begin{equation} \label{eq__35}
\small{B=\frac{32}{\pi }\int\limits_{\sqrt{1-{{b}^{2}}}}^{1}{dx}\int\limits_{\sqrt{1-{{b}^{2}}}}^{1}{dy\frac{xD\left( {{\alpha }_{2}}=x \right)D\left( {{\alpha }_{2}}=y \right)K(k)}{{{(n+m)}^{2}}}},\ y>x.}
\end{equation}
Here
\[\begin{matrix}
  K(k)=\int\limits_{0}^{{\pi }/{2}\;}{\frac{d\varphi }{\sqrt{1-{{k}^{2}}{{\sin }^{2}}\varphi }},\quad k={{\left( \frac{n-m}{n+m} \right)}^{2}};} \\
  n={{\left( 1-{{x}^{2}} \right)}^{{1}/{4}\;}}+{{\left( 1-\frac{{{x}^{2}}}{{{y}^{2}}} \right)}^{{1}/{4}\;}};\  \\
  m=8\left\{ {{\left[ \left( 1-{{x}^{2}} \right)\left( 1-\frac{{{x}^{2}}}{{{y}^{2}}} \right) \right]}^{{1}/{4}\;}} \right.\times  \\
  \times {{\left. \left[ {{\left( 1-{{x}^{2}} \right)}^{{1}/{2}\;}}+{{\left( 1-\frac{{{x}^{2}}}{{{y}^{2}}} \right)}^{{1}/{2}\;}} \right] \right\}}^{{1}/{4}\;}} \\
\end{matrix}\].
For $x>y$ symmetric integrands are obtained, in which x and y are interchanged. For this reason it is sufficient to calculate the integrals for $x>y$ and multiply the result obtained by 2. In the formulas presented this multiplication has already been performed. Since the series $K(k)$ converges very rapidly, for the calculations we can take $K(k)={\pi }/{2}\;$. A somewhat different form of $B$, obtained by substitution of variables, is more convenient for calculations:
\begin{equation} \label{eq__36}
\begin{matrix}
  B=\frac{32{{b}^{3}}}{\pi }\int\limits_{0}^{1}{dx}\int\limits_{0}^{1}{dy\frac{xyD\left( {{\alpha }_{1}}=x \right)D\left( {{\alpha }_{1}}=y \right)K({{k}_{1}})}{{{\left( {{y}^{2}}-{{b}^{-2}}-1 \right)}^{{1}/{2}\;}}{{({{n}_{1}}+{{m}_{1}})}^{2}}}},\ y>x; \\
  {{k}_{1}}={{\left( \frac{{{n}_{1}}-{{m}_{1}}}{{{n}_{1}}+{{m}_{1}}} \right)}^{2}}; \\
  {{n}_{1}}={{\left[ {{b}^{2}}\left( 1-{{x}^{2}} \right) \right]}^{{1}/{4}\;}}+{{\left( \frac{{{y}^{2}}-{{x}^{2}}}{{{y}^{2}}+{{b}^{-2}}+1} \right)}^{{1}/{4}\;}}; \\
  {{m}_{1}}=\left\{ 8{{\left[ \frac{{{b}^{2}}\left( 1-{{x}^{2}} \right)\left( {{y}^{2}}-{{x}^{2}} \right)}{{{y}^{2}}-{{b}^{-2}}-1} \right]}^{{1}/{4}\;}} \right.\times  \\
  {{\left. \times \left[ {{\left( {{b}^{2}}\left( 1-{{x}^{2}} \right) \right)}^{{1}/{2}\;}}+{{\left( \frac{{{y}^{2}}-{{x}^{2}}}{{{y}^{2}}+{{b}^{-2}}-1} \right)}^{{1}/{2}\;}} \right] \right\}}^{^{^{{1}/{4}\;}}}}. \\
\end{matrix}
\end{equation}
Substituting the numerical values into \eqref{eq__34} and \eqref{eq__36} and integrating, we obtain
\begin{equation} \label{eq__37}
A=0.088;\quad B=0.0137.
\end{equation}
The averaged form factors for a heterocontact in the model of a clean opening satisfy the expression
\begin{equation} \label{eq__38}
{\left\langle {{K}_{Ta}} \right\rangle }/{\left\langle {{K}_{0}} \right\rangle }\;=A\left\langle {{\alpha }_{1}}D\left( {{\alpha }_{1}} \right) \right\rangle _{{{v}_{z}}>0}^{-1}=0.296;
\end{equation}
\begin{equation} \label{eq__39}
\begin{matrix}
  {\left\langle {{K}_{Cu}} \right\rangle }/{\left\langle {{K}_{0}} \right\rangle }\;=B\left\langle {{\alpha }_{2}}D\left( {{\alpha }_{2}} \right) \right\rangle _{{{v}_{z}}>0}^{-1}= \\
  =B{{b}^{-2}}\left\langle {{\alpha }_{1}}D\left( {{\alpha }_{1}} \right) \right\rangle _{{{v}_{z}}>0}^{-1}=0.150. \\
\end{matrix}
\end{equation}
Since the contribution of copper to the spectrum of the heterocontact is not noticeable, it may be assumed without introducing any special error that the entire spectrum is determined by the tantalum edge. In this case the more accurate value of the intensity of the PC EPI function of tantalum in the hererocontact equals
\begin{equation} \label{eq__40}
g_{pc}^{het}={{A}^{-1}}\left\langle {{\alpha }_{1}}D\left( {{\alpha }_{1}} \right) \right\rangle \left( {{{d}_{Ta}}}/{{{d}_{het}}}\; \right){{g}_{pc}},
\end{equation}
Where ${g}_{pc}$ is the PC EPI function of the $Ta-Ta$ homocontact with the same resistance (see \eqref{eq__3} and \eqref{eq__4}). From here it is necessary to calculate correction factor
\begin{equation} \label{eq__41}
\begin{matrix}
  {{k}_{korr}}=\lambda _{pc}^{het}/\lambda _{pc}^{f}=\ g_{pc}^{het}/g_{pc}^{f}=\  \\
  =\sqrt{2}{{A}^{-1}}\left( v_{F}^{e}k_{F}^{e}/v_{F}^{f}k_{F}^{f}\  \right)\left\langle {{\alpha }_{1}}D\left( {{\alpha }_{1}} \right) \right\rangle _{{{v}_{z}}>0}^{3/2\;}=0.674 \\
\end{matrix}
\end{equation}
Here $g_{pc}^{f}$ and $\lambda _{pc}^{f}$ are the absolute intensity of the PC EPI function and the parameter $\lambda $, calculated in the free-electron approximation for a $Ta-Ta$ homocontact with the same resistance.

We shall calculate the ratio of the averaged form factors for a heterocontact in the ballistic regime. As follows from the formulas \eqref{eq__38} and \eqref{eq__39},
\begin{equation} \label{eq__42}
{\left\langle {{K}_{Cu}} \right\rangle }/{\left\langle {{K}_{Ta}} \right\rangle =B/{{b}^{2}}A}\;=0.506.
\end{equation}
This quantity is greater than ${{b}^{2}}=0.307$, which is presented for a dirty heterocontact\footnote {According to a remark by R.I. Shekhter, this is attributable to the strong angular dependence of the form factors, which accentuate the scattering at large angles. In the presence of an atomically smooth boundary between the mewls, ${\left\langle {{K}_{Cu}} \right\rangle }/{\left\langle {{K}_{Ta}} \right\rangle }\;\simeq b$. If the transit regime of the electrons moving through the point contact remains ballistic, but the boundary between the metals is rough and scatters electrons diffusely, then ${\left\langle {{K}_{Cu}} \right\rangle }/{\left\langle {{K}_{Ta}} \right\rangle }\;\simeq {{b}^{2}}$.}
From here the partial contributions of $C$u and $Ta$ to the spectrum in the ballistic regime can be determined more accurately (compare with \eqref{eq__7}):
\begin{equation} \label{eq__43}
L=v_{F}^{Ta}g_{pc}^{Cu}{\left\langle {{K}_{Cu}} \right\rangle }/{v_{F}^{Cu}}\;g_{pc}^{Ta}\left\langle {{K}_{Ta}} \right\rangle .
\end{equation}
Taking this into account we can put the formula \eqref{eq__16} into a different form:
\begin{equation} \label{eq__44}
\frac{{{L}_{f}}}{{{L}_{e}}}={{\left( \frac{k_{F}^{Cu}}{k_{F}^{Ta}} \right)}_{f}}{{\left( \frac{k_{F}^{Ta}}{k_{F}^{Cu}} \right)}_{e}}{{\left( \frac{\left\langle {{K}_{Cu}} \right\rangle }{\left\langle {{K}_{Ta}} \right\rangle } \right)}_{f}}{{\left( \frac{\left\langle {{K}_{Ta}} \right\rangle }{\left\langle {{K}_{Cu}} \right\rangle } \right)}_{e}}\simeq 1.041.
\end{equation}
Table I shows the parameters of homo- and heterocontacts in the free-electron approximation and for two values of the Fermi velocity of tantalum.
\begin{table*}[]
\caption[]{}
\begin{tabular}{|p{1in}|p{1.3in}|p{1.3in}|p{1.5in}|} \hline
Parameter & $v_{F} =1.369\cdot 10^{8} \, cm/s$; $k_{F} =1.183\cdot 10^{8} \, cm/s$ & $v_{F} =0.51\cdot 10^{8} \, cm/s$; $k_{F} =0.82\cdot 10^{8} \, cm/s$ & $v_{F} =0.24\cdot 10^{8} \, cm/s$ \cite{Shen1};\newline $k_{F} =0.348\cdot10^{8} \, cm/s$ \\ \hline
b & 0.872 & 0.554 & 0.259 \\ \hline
c & 0.872 & 0.325 & 0.153 \\ \hline
$\left\langle \alpha _{1} D\left(\alpha _{1} \right)\right\rangle _{v_{z} >0} $ & 0.479 & 0.299 & 0.168 \\ \hline
${\left\langle K_{Ta} \right\rangle \mathord{\left/ {\vphantom {\left\langle K_{Ta} \right\rangle  \left\langle K_{0} \right\rangle }} \right. \kern-\nulldelimiterspace} \left\langle K_{0} \right\rangle } $ & 0.468 & 0.296 & 0.168 \\ \hline
${\left\langle K_{Cu} \right\rangle \mathord{\left/ {\vphantom {\left\langle K_{Cu} \right\rangle  \left\langle K_{0} \right\rangle }} \right. \kern-\nulldelimiterspace} \left\langle K_{0} \right\rangle } $ & 0.387 & 0.150 & 0.039 \\ \hline
${\left\langle K_{Cu} \right\rangle \mathord{\left/ {\vphantom {\left\langle K_{Cu} \right\rangle  \left\langle K_{Ta} \right\rangle }} \right. \kern-\nulldelimiterspace} \left\langle K_{Ta} \right\rangle } $ & 0.827 & 0.506 & 0.234 \\ \hline
$d_{{\rm homo}} ,nm$ & $37.6\left(R\left[\Omega \right]\right)^{{-1\mathord{\left/ {\vphantom {-1 2}} \right. \kern-\nulldelimiterspace} 2} } $ & $54.26\left(R\left[\Omega \right]\right)^{{-1\mathord{\left/ {\vphantom {-1 2}} \right. \kern-\nulldelimiterspace} 2} } $ & $115.86\left(R\left[\Omega \right]\right)^{{-1\mathord{\left/ {\vphantom {-1 2}} \right. \kern-\nulldelimiterspace} 2} } $ \\ \hline
$d_{{\rm het}} ,nm$ & $38.4\left(R\left[\Omega \right]\right)^{{-1\mathord{\left/ {\vphantom {-1 2}} \right. \kern-\nulldelimiterspace} 2} } $ & $70.2\left(R\left[\Omega \right]\right)^{{-1\mathord{\left/ {\vphantom {-1 2}} \right. \kern-\nulldelimiterspace} 2} } $ & $200.1\left(R\left[\Omega \right]\right)^{{-1\mathord{\left/ {\vphantom {-1 2}} \right. \kern-\nulldelimiterspace} 2} } $ \\ \hline
$k_{n} $ & 1 & 3.87 & 17.57 \\ \hline
$g_{pc}^{\max } $ & 0.622 & 0.16 & 0.035 \\ \hline
$k_{korr} $ & 2.088 & 0.674 & 0.196 \\ \hline
$g_{pc\, \max }^{het} $ & 1.157 & 0.374 & 0.109 \\ \hline
$L$ & 0.262 & 0.269 & 0.268 \\ \hline
$L\left(g_{pc\, \max }^{het} \right)$ & 0.14 & 0.115 & 0.086 \\ \hline
$\lambda _{pc}^{{\rm homo}} $ & 0.88 & 0.227 & 0.05 \\ \hline
$\lambda _{pc}^{{\rm het}} $ & 1.71 & 0.553 & 0.161 \\ \hline
$\lambda _{tr} $ & 0.854 & 0.456 & 0.215 \\ \hline
$J$ & 1.236 & 0.575 & 0.259 \\ \hline
\end{tabular}
\label{Table1}
\end{table*}
The value $v_{F}^{Ta}=0.2\text{4}\cdot \text{1}{{0}^{8}}cm/\sec $ is taken from Ref. \cite{Shen1}. The corresponding wave vector ${{k}_{F}}=0.38\text{4}\cdot \text{1}{{0}^{8}}cm/\sec $ is determined analogously to the preceding wave vector (see formula \eqref{eq__12}) using ${m^*}/{{{m}_{0}}=1.85}\;$ from Ref. \cite{Halloran}. The expected contribution of copper to the spectrum of the heterocontact was evaluated using the formula \eqref{eq__43}. In this case the values of $L$ corresponded to the use of the absolute intensity of the PC EPI function, obtained for $Ta-Ta$ homocontacts, and the values $L\left( g_{pc}^{het} \right)$ corresponded to the use of the more accurate absolute intensity of the PC EPI function of tantalum in $Ta-Cu$ heterocontacts. Although it follows from Table I that the contribution of copper to the spectrum of the heterocontact in the estimates must exceed 25\%, the contribution of copper is not noticeable in the spectra of heterocontacts. At the same time, if the absolute intensity of the PC EPI function ${{g}_{pc}}$ and the parameter ${{\lambda }_{pc}}$ for tantalum in the $Ta-Cu$ heterocontacts are calculated just as for $Ta-Ta$ homocontacts with the same resistance, we obtain values that are close to those obtained for homocontacts. Thus, for the best heterocontact in the free-electron approximation, $g_{pc}^{\max f}=0.\text{554}$ and $\lambda _{pc}^{f}=0.\text{82}$. From here there follow two possibilities: 1) the maximum intensity of the spectrum is not reached in the homocontacts; 2) in the existing technology of forming the contact, an unsymmetric heterocontact is formed, and tantalum occupies most of the effective volume of phonon generation near the constriction. We shall study both possibilities separately. In the analysis we shall assume that $v_{F}^{Ta}=0.\text{51}\cdot \text{1}{{0}^{\text{8}}}\text{cm}/\text{sec}$.
\section{SYMMETRIC CONTACT}
\begin{figure}[]
\includegraphics[width=8cm,angle=0]{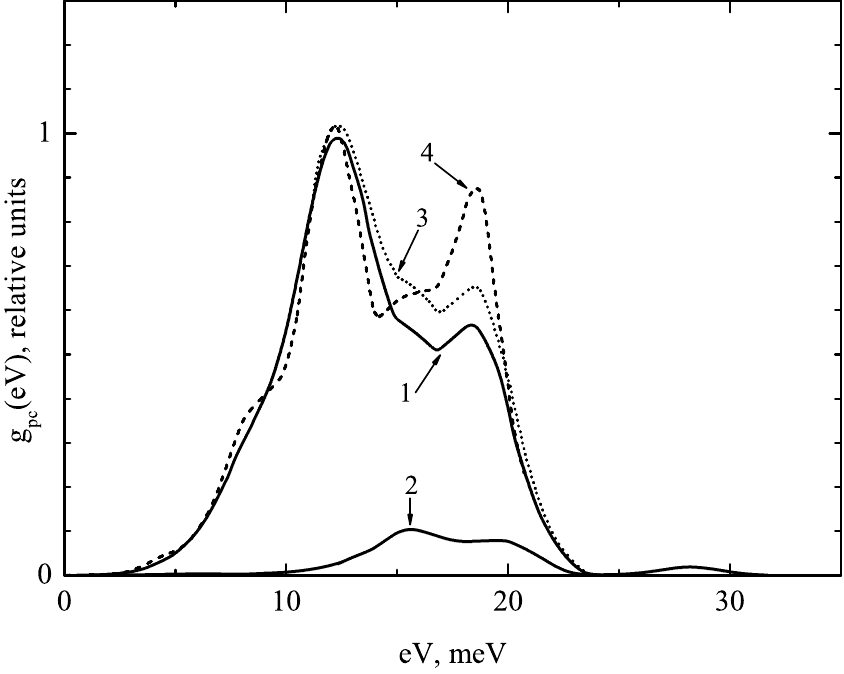}
\caption[]{Comparison of the PC EPI functions of $Ta$ in homo- and heterocontacts: \\1) $Ta-Ta$, curve 2 from Fig. \ref{Fig2}; \\2) PC EPI function of $Cu$ from Ref. \cite{Yanson2}; \\3) the summed curve $g_{p{{c}_{_{^{Cu}}}}}^{\max }=0.1g_{p{{c}_{Ta}}}^{\max }$; \\4) $Ta-Cu$, curve 6 from Fig.\ref{Fig2}.}
\label{Fig3}
\end{figure}

In this case the maximum corrected intensity of the PC EPI and the relative contribution of copper to the spectrum $g_{pc\max }^{het}=0.\text{374}$, copper relative contribution to the spectrum  $L\left( g_{pc}^{het} \right)=0.115$, i.e. it is only somewhat greater than 10\% and indeed should not be noticeable. Figure \ref{Fig3} shows an example of the addition of the PC EPI functions of tantalum and copper homocontacts with the relative intensity $g_{pc\max }^{Cu}=0.1g_{pc\max }^{Ta}$.

As one can see the contribution of copper to the total spectrum is not noticeable, though the form of the total spectrum differs from the spectrum of the heterocontact - in the case of heterocontacts the relative intensity of the $L$ peak is almost 40\% higher.

One possible reason for this strong difference in absolute intensities and the form of Ta spectra in homo- and heterocontacts could be the following.

The difference in the absolute intensities of PC EPI functions and Eliashberg's thermodynamic EPI function is linked with the different form factors - in the case of the thermodynamic function every electron scattering event is effective, and for the PC EPI function large-angle scattering processes make the main contribution. This is usually also linked with the decrease in the relative intensity of the high-frequency maximum in the PC spectra \cite{Lee}. In a heterocontact the tangential component of the electron momentum is conserved across the interface. For this reason any electron that has undergone inelastic scattering in tantalum and whose $z$ component of the momentum has been reversed will return into the copper, being deflected after refraction at the boundary toward the $z$ axis, which increases the contribution of small-angle scattering processes in $Ta$ to the "return current", which determines the inelastic correction to the current through the point contact. (We note that the maximum angle of incidence of the electrons from the copper side, when total internal reflection does not yet occur, equals ${{33}^{\circ }}$ with respect to the $z$ axis.) At the same time the form factor for the metal with the lower value of ${{p}_{F}}(Ta)$ also approaches the form factor of Eliashberg's thermodynamic function, while the absolute intensities of the PC function ${{g}_{pc}}$ and the EPI parameter ${{\lambda }_{pc}}$ become closer to the corresponding values of the thermodynamic function. This is also indicated by the closer form of the PC EPI function of the heterocontact (Fig. \ref{Fig2}, curves 4-6) and the thermodynamic EPI function (Fig. \ref{Fig2}, curve 7).

We note that the PC spectrum of the metal with the large value of ${{p}_{F}}$ in the heterocontact undergoes opposite changes, owing to the fact that the scattering by angles $\theta <{{\theta }_{\min }}$ does not contribute to the return current (see Fig. \ref{Fig2} in Ref. \cite{Baranger}) and thereby accentuates the effect of the $K$ factor, which emphasizes the scattering by large angles. Therefore if we could separate out the low intensity contribution to the PC spectrum from the metal with large $p_F$, we would see that its form differs even more strongly from Eliashberg's EPI function than the form of the PC EPI function obtained from the spectrum of the homocontact of this metal. At the same time, as one can see from Table I, a physically more reliable result for $\lambda _{pc}^{het}$ is obtained using $v_{F}^{Ta}=0.\text{51}\cdot \text{1}{{0}^{\text{8}}}\text{cm}/\text{sec}$.
\section{UNSYMMETRIC HETEROCONTACT}
The magnitude of the excess current when there is no geometric asymmetry and additional scatterers at the boundary in a clean heterocontact of the $S-c-N$ type must equal \cite{Zaitsev}
\begin{equation} \label{eq__45}
\small{{I}_{exc}}=\frac{\Delta }{{{R}_{N}}}J,\ J=\frac{1}{2\left\langle \alpha D \right\rangle }{{\left\langle \alpha \frac{{{D}^{2}}}{R}\left[ 1-\frac{{{D}^{2}}arth\sqrt{R}}{\sqrt{R}(1+R)} \right] \right\rangle }_{{{v}_{z}}>0}},
\end{equation}
where $R=1-D$. Substituting $D\left( {{\alpha }_{1}} \right)$ into the formula for $J$ and integrating the expression obtained we obtain $J=0.575$. Thus for $Ta-Cu$ symmetric heterocontacts ${{I}_{exc}}=0.575\,{\Delta }/{{{R}_{N}}}\;$. In the limit $D\to 1$ we have $J=\text{4}/\text{3}$. For clean $S-c-S$ homocontacts in the presence of a semitransparent barrier between the metals D does not depend on $\alpha$. In this case
\begin{equation} \label{eq__46}
{{I}_{exc}}=\frac{2\Delta }{{{R}_{N}}}J,\ J=\frac{D}{2R}\left[ 1-\frac{{{D}^{2}}arth\sqrt{R}}{\sqrt{R}(1+R)} \right].
\end{equation}
The deviation of $D$ from 1 affects the magnitude of the excess current much more strongly than it affects the intensity of the spectrum. Thus the excess current for $D=0.7$ is half of the excess current for $D=1$, while the intensity of the spectrum in this case equals ${{D}^{{1}/{2}\;}}=0.84$ of the starting intensity.

In our experiments there was no correlation between the absolute intensity of the spectrum and the magnitude of the excess current, observed in Refs. \cite{Yanson3} and \cite{Khotkevich} for tin contacts. The superconducting characteristics turned out to be more sensitive to contamination than the spectral characteristics. The magnitude of the excess current in the contacts with an intense spectrum in the N state could vary from 0 up to 0.8 of the theoretical value for a clean $S-c-S$ contact, but the critical Josephson current was almost always absent. This indicates the presence of depairing centers in the plane of the constriction. In heterocontacts with an intense PC spectrum the parameter $J$ could vary from 0 to 1.1. From here it follows that at least for $J\gtrsim 0.6$ a geometric asymmetry of the heterocontact existed and $Ta$ filled most of the volume near the constriction, and in addition in this asymmetric heterocontact for $J$ close to 1.1 the ballistic regime was realized.

In this case it is obvious that the absolute intensity of the PC function ${{g}_{pc}}$ and the EPI parameter ${{\lambda }_{pc}}$, calculated for the $Ta-Cu$ heterocontact, are close, just like for a $Ta-Ta$ heterocontact with the same resistance and the same characteristics calculated for $Ta-Ta$ homocontacts. The reason for the fact that copper does not contribute to the spectrum is also understandable. The decrease in the relative intensity of the $L$ peak in the homocontact compared with the heterocontacts could be linked with the heating of the contact for energies close to the Debye energies. The typical resistances of our contacts equal $20-100\  \Omega$, which corresponds to diameters of 120-50 {\AA}. If the energy mean-free path length is evaluated from the relation ${{l}_{\varepsilon }}={{v}_{F}}{{\tau }_{\varepsilon }}$ (here $\tau _{\varepsilon }^{-1}=2\pi {{\hbar }^{-1}}\int\limits_{0}^{\varepsilon }{g(\omega )d\omega }$) is the energy relaxation time and $g(\omega )$ is Eliashberg's thermodynamic function, given, e.g., in the tables of Ref. \cite{Rowell} then for $\omega ={{\omega }_{D}}$ we have ${{l}_{\varepsilon }}=146{\AA}$, i.e., contacts with the lowest resistances at energies close to the Debye energies must be in a state close to the thermal state, when spectroscopy is impossible. At the same time, although the background for the high-resistance contacts is high (50-70\%) and the intensity of the $L$ peak is lowered appreciably, it is nonetheless quite well resolved. This indicates that it is incorrect to use the relation indicated above for determining the energy relaxation time, since this relation is predicated on the equilibrium electron distribution function. A substantially nonequilibrium electron distribution function is realized in the point contact near the constriction, and although the energy electron mean-free path length is much greater than the diameter of the contact, it is still not large enough, which increases the background and decreases the intensity of the $L$ peak. It is much easier to satisfy the condition ${{l}_{\varepsilon }}\gg d$ with the use of copper as the counterelectrode, since there are no restrictions from the side of the copper edge and the condition ${{l}_{\varepsilon }}\gg r$ ($r=d/2$ is the radius of the contact) must be satisfied. The decrease in the background level in heterocontacts is also explained by the lower heating of the contact owing to the much better thermal conductivity of the copper edge, which efficiently removes heat.

\begin{figure}[t]
\includegraphics[width=8cm,angle=0]{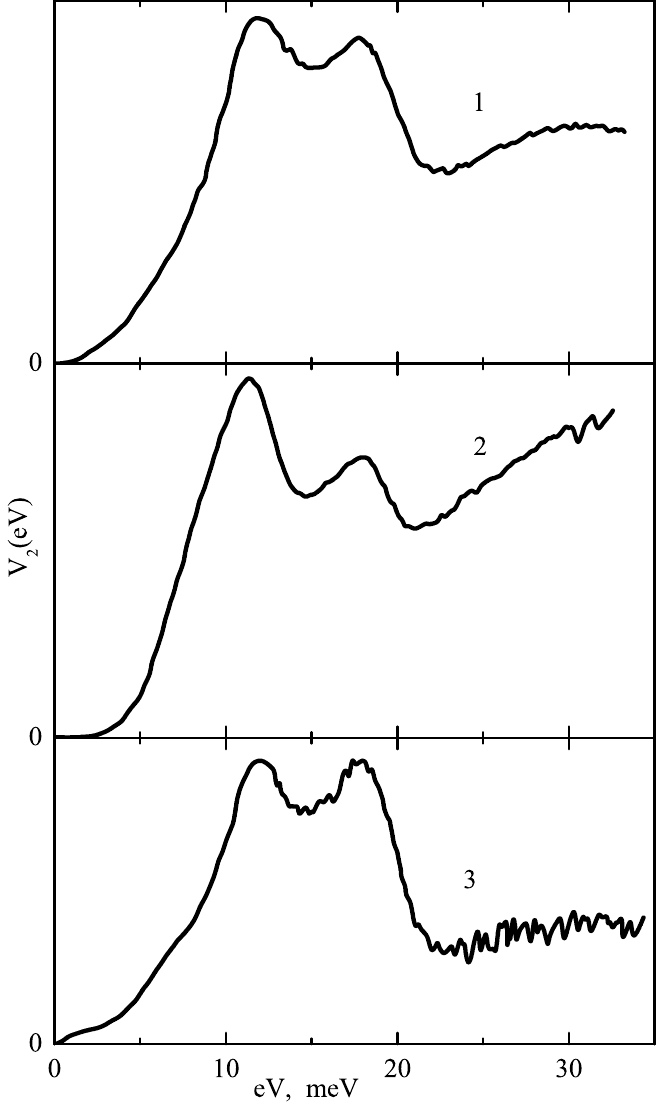}
\caption[]{Check of the effect of the technology employed to fabricate the point contact on the form of the PC EPI spectrum ${{V}_{2}}\sim{\ }{dV}/{d{{I}^{2}}}\;$, for electroforming (1) and using the shear method (2, 3): \\1) $Ta-Au$, $R=70\ \Omega $, ${{V}_{1}}(0)=710\ \mu V$, $v_{2}^{\max }=0.6\,\mu V$, $T=4.8\,K,\ H=0$;
\\2) $Ta-Au$, $R=5\ \Omega $, ${{V}_{1}}(0)=545\ \mu V$, $v_{2}^{\max }=0.57\,\mu V$, $T=5\,K,\ H=0$;
\\3) $Ta-Cu$, $R=30\ \Omega $, ${{V}_{1}}(0)=754\ \mu V$, $v_{2}^{\max }=0.54\,\mu V$, $T=4.8\,K,\ H=0$.}
\label{Fig4}
\end{figure}

The two possibilities, however, are not mutually exclusive. Both symmetric and unsymmetric heterocontacts with a dominant contribution of tantalum are apparently realized. At the same time, as the interface between the metals moves away from the plane of the geometric symmetry of the contact, the effective volume of phonon generation in tantalum increases and the effect of the $\delta $ function barrier between the metals owing to the difference in the Fermi velocities decreases. At the same time, the effect of the form factor, which is determined by the difference in the Fermi momenta and the intensity of the PC EPI function of the metal with the smaller value of ${{\rho }_{F}}$, approaching Eliashberg's thermodynamic function, also decreases. In $Ta-Cu$ heterocontacts, the mutual effect of these factors is largely compensated, and for this reason the contribution of copper to the spectrum is not noticeable, while the intensity of the spectra does not change much as the symmetry of the contact changes. The magnitude of the excess current, however, changes at the same time over a wide range. To check this proposition, we prepared a series of $Ta-Ta$, $Ta-Cu$, and $Ta-Au$ contacts using the standard shear method, with which a geometrically symmetric contact is most likely to be formed. As in the case of the electroforming of contacts, however, no contribution from copper and gold to the spectrum was observed in heterocontacts. The absolute intensity of the spectra for $Ta-Ta$ homocontacts and $Ta-Cu$ heterocontacts in this case was much lower than with electroforming, and turned out to be close to the absolute intensity of the spectra which we obtained previously \cite{Rybal'chenko}. In the case of $Ta-Au$ point contacts, the absolute intensity of the PC spectra obtained for electrically formed point contacts turned out to be only slightly higher than for point contacts prepared by the standard shear method \cite{Yanson1}. This is apparently explained by the mutual solubility of $T$a and $Au$.
Figure \ref{Fig4} shows typical PC spectra of $Ta-Cu$ and $Ta-Au$ for contacts prepared by different methods.
All values of $g_{pc}^{f}$ and $\lambda _{pc}^{f}$ in the caption in Fig. \ref{Fig2} correspond to the free-electron approximation, and in addition in heterocontacts the values of $g_{pc}^{f}$ and $\lambda _{pc}^{f}$ were calculated just as for homocontacts with the same resistance.
All estimates made above must be treated with care. Since $d$ metals have two groups of valence electrons, in different experiments one or another group of electrons can make the dominant contribution. For example, in niobium, which is the electronic analog of tantalum, the Fermi velocity, determined in optical measurements, $v_{F}^{Nb}=0.9\text{4}\cdot \text{1}{{0}^{8}}cm/\sec $ (Ref. \cite{Mash}). In Ref. \cite{Blonder}, in order to make the computational results agree with experiment for $Cu-Nb$ point contacts, the value $v_{F}^{Nb}=0.897\cdot \text{1}{{0}^{8}}cm/\sec $ was employed. For different values of ${{v}_{F}}$ and ${{p}_{F}}$ the values of $J$ and $D$ will be different.
\section{CONCLUSIONS}
In conclusion we shall list the basic results obtained in this work.
 A new procedure was developed for preparing clean point contacts.
\begin{enumerate}
\item {The EPI parameter ${{\lambda }_{pc}}$ and the absolute intensity of the PC function ${{g}_{pc}}(\omega )$ for tantalum were determined, and the exact form of the PC function was found.}
\item {The PC characteristics of a symmetric heterocontact were calculated numerically for the first time in this work: the absolute intensity of the EPI function ${{g}_{pc}}(\omega )$ and the parameter ${{\lambda }_{pc}}$. At the same time, an expression taking into account the degree to which the electronic parameters of each electrode affect the form factor of the heterocontact was derived, which makes it possible to evaluate the relative contribution of the specific metal to the resulting spectrum. The calculation is based on the results of the theory of Shekhter and Kulik \cite{Shekhter}, and the case of the ballistic contacts in the approximation of a spherical Fermi surface was examined.}
\item {It was established that the contribution of the second metal to the PC spectra of the heterocontacts $Ta-Cu$ and $Ta-Au$ is not noticeable, while the intensities of the PC spectra remain virtually constant as the geometric symmetry of the contacts changes. At the same time the forms of the spectra in homo- and heterocontacts are different. An explanation of these phenomena with the help of the compensation effect was proposed.}
\item {The effective mass approximation was used for the first time to calculate the PC characteristics, since the free-electron approximation obviously leads to incorrect results.}
\item {A new procedure was developed for preparing clean point contacts.}
\end{enumerate}
We thank V.M. Kirzhner, O.I. Shklyarevskii, and V.V.~Khotkevich for assistance in the computer calculations.
\section{NOTATION}
Here ${{v}_{F}}$ and ${{p}_{F}}$ are the Fermi velocity and Fermi momentum, respectively; $g$ and $\lambda $ are the EPI function and parameter, respectively; ${{k}_{F}}$ is the Fermi wave vector; $d$ is the diameter of the contact; $a$ is the lattice constant; $\Omega $ is the volume of the unit cell; $L$ is the relative intensity of the partial spectra of a heterocontact; $\gamma $ is the electronic heat capacity; and, $D$ is the coefficient of electron transit through the boundary in a heterocontact.


\begin{thebibliography}{}
\bibitem{Yanson1} I.K. Yanson. \href{http://fntr.ilt.kharkov.ua/fnt/pdf/9/9-7/f09-0676r.pdf}{Fiz.Nizk. Temp.} 9, 676 (1983) [Sov. J. Low Temp. Phys \textbf{9}, 343 (1983)].

\bibitem{Rybal'chenko} L.F. Rybal'chenko, I.K. Yanson, N.L. Bobrov, and V.V. Fisun, \href{http://fntr.ilt.kharkov.ua/fnt/pdf/7/7-2/f07-0169r.pdf}{Fiz. Nizk. Temp.} \textbf{7}, 169 (1981) [Sov. J. Low Temp. Phys. 7, 82 (1981)].

\bibitem{Ryazanov} V.V. Ryazanov, V.V. Schmidt, and L.A. Ermolaeva, \href{http://link.springer.com/article/10.1007/BF00654497}{J. Low Temp.Phys.}\textbf{45}, No. 5/6, 507 (1981).

\bibitem{Startsev} V.E. Startsev, "Local singularities in the Fermi surfaces and electronic transport phenomena in transition metals" Author's Abstract of Doctoral Dissertation in Physical-Mathematical Sciences, Sverdlovsk (1983)

\bibitem{Bostok} K.J. Bostok, W.N. Cheung, R.M. Rose, and M.L.A. MacVicar, Inst. Phys. Conf. Ser. \textbf{8}, No. 39, 662 (1978).

\bibitem{Engen} Engen, Aies, Adams, Fombarlet, Pribory dlya nauchnykh issledovanii, No.12. 132 (1989), [\href{http://dx.doi.org/10.1063/1.1137965}{Rev. Sci. Instrum.} 55, 1489 (1984)]

\bibitem{Kulik} I.O. Kulik, A.N. Omel'yanchuk, and R.I. Shekhter, Fiz.Nizk. Temp. \textbf{3}, 1543 (1977) [Sov. J. Low Temp. Phys. \textbf{3}, 740 (1977)].

\bibitem{Harrison} W. Harrison, \href{http://www.amazon.com/Solid-State-Theory-Dover-Physics/dp/0486639487}{Solid-State Theory}, Dover (1980).

\bibitem{Ajhcroft} N.W. Ajhcroft and N.D. Mermin, \href{http://www.worldcat.org/title/solid-state-physics/oclc/934604}{Solid State Physics}, Holt, Rinehart, and Winston (1976).

\bibitem{Wolf} E.L. Wolf, D. M. Burnell, Z.G. Klein, and R.J. Noert, \href{http://link.springer.com/article/10.1007/BF00115078}{J. Low Temp.Phys.} \textbf{44}, No. 1/2, 89 (1981).

\bibitem{Pishche} Progress in the Electronic Theory of Metals [Russian translation, edited by P. Pishche and P. Lemman], Mir, Moscow (1984), Vol. 2.

\bibitem{Shekhter} R.I. Shekhter and I.O. Kulik, \href{http://fntr.ilt.kharkov.ua/fnt/pdf/9/9-1/f09-0046r.pdf}{Fiz. Nizk. Temp.} \textbf{9}, 46 (1983) [Sov. J. Lov,Temp. Phys. 9, 22 (1983)].

\bibitem{Yanson2} I.K. Yanson, A.I. Akimenko, and A. B. Verkin, \href{http://www.sciencedirect.com/science/article/pii/0038109882909887}{Solid State Commun.} 43, No.10, 765 (1982).

\bibitem{Mattheiss} L.F. Mattheiss, \href{https://journals.aps.org/prb/abstract/10.1103/PhysRevB.1.373}{Phys. Rev. B} \textbf{1}, No. 2, 373 (1970).

\bibitem{Alekseevskii} N.E. Alekseevskii, V.I. Nizhankovskii, and K.-Kh. Bertel', \href{http://impo.imp.uran.ru/fmm/Electron/vol37_1/abstract8.pdf}{Fiz. Met.Metalloved} \textbf{37}, No. 1, 63 (1974).

\bibitem{Halloran} M. H. Halloran, J. H. Condon, J. E. Graebner, J. E. Kunzier, and F. S. L. Hsu, \href{http://journals.aps.org/prb/abstract/10.1103/PhysRevB.1.366}{Phys. Rev. B} \textbf{1}, No. 2. 366 (1970).

\bibitem{Gniwek} J.J. Gniwek, J.C. Moulder, and R.H. Kropscot. "Electrical conductivity of high purity copper" in: Proceedings of the \href{https://books.google.com.ua/books?id=j1teOwAACAAJ&dq=bibliogroup:%22Proceedings+of+the+...+International+Conference+on+Low+Temperature+Physics+;+LT-10%22&hl=ru&sa=X&ved=0ahUKEwixgM_ys77JAhWHGCwKHR3BC3MQ6AEIGzAA}{10th International Conference on Low Temperature Physics}, VINITI, Moscow (1967), Vol. 3. pp. 336-370.

\bibitem{Shen1} L.Y.L. Shen, "Superconductivity of tantalum, niobium and lanthanum studied by electron tunneling problems of surface contamination" in: \href{http://www.researchgate.net/publication/234851839_Superconductivity_of_Tantalum_Niobium_and_Lanthanum_Studied_by_Electron_Tunneling_Problems_of_Surface_Contamination}{Superconductivity in d- and f-Band Metals.} edited by D.H. Douglass, Plenum Press, New York (1972), pp. 31-44.

\bibitem{Lee} M.J.G. Lee, J. Caro, O.G. Croot, and R. Griessen. \href{http://journals.aps.org/prb/abstract/10.1103/PhysRevB.31.8244#fulltext#fulltext}{Phys. Rev. B} \textbf{31}, No. 12, 8244 (1985).

\bibitem{Baranger} H.U. Baranger, A.H. MacDonald, and C.R. Leavens, \href{http://journals.aps.org/prb/abstract/10.1103/PhysRevB.31.6197#fulltext#fulltext}{Phys. Rev. B} \textbf{31}, No. 10, 6197 (1985).

\bibitem{Zaitsev} A.V. Zaitsev, Zh. Eksp. Teor.Fiz.\textbf{86}, 1742 (1984) [\href{http://www.jetp.ac.ru/cgi-bin/dn/e_059_05_1015.pdf}{Sov.Phys.JETP} \textbf{59}, 1015 (1984)].

\bibitem{Yanson3} I.K. Yanson, G.V. Kamarchuk, and A.V. Khotkevich, \href{http://fntr.ilt.kharkov.ua/fnt/pdf/10/10-4/f10-0423r.pdf}{Fiz. Nizk. Temp.}\textbf{10}, 423 (1984). [Sov. J. Low Temp. Phys. \textbf{10}, 220 (1984)].

\bibitem{Khotkevich} A.V. Khotkevich and I.K. Yanson. \href{http://fntr.ilt.kharkov.ua/fnt/pdf/7/7-6/f07-0727r.pdf}{Fiz.Nizk Temp.} \textbf{7}, 727 (1981) [Sov.J. Low Temp. Phys. \textbf{7}, 354 (1981)].

\bibitem{Rowell} J.M. Rowell, W.L. McMillan, and R.C. Dynes, "A tabulation of the electron-phonon interaction in superconducting metals and alloys. Part I." Preprint, Bell Lab., Munay Hill (1973).

\bibitem{Mash} I.D. Mash, Tr. Fiz. in-ta im. P. N. Lebedev, Akad. Nauk SSSR, \textbf{82}, 2 (1975).

\bibitem{Blonder} G.E. Blonder and M. Tinkham, \href{http://journals.aps.org/prb/abstract/10.1103/PhysRevB.27.112#fulltext#fulltext}{Phys. Rev. B} \textbf{27}, No. 1, 112 (1983).

\bibitem{Shen2} L.Y.L. Shen, \href{http://journals.aps.org/prl/abstract/10.1103/PhysRevLett.24.1104#fulltext#fulltext}{Phys. Rev. Lett} \textbf{24}, No. 20, 1104 (1970).

\end{thebibliography}
\end{document}